\documentclass[12pt]{article}
\usepackage{amssymb,amsmath}
\begin{document}

\begin{titlepage}

                            \begin{center}
                            \vspace*{1cm}
\Large\bf{Long-range interactions, doubling measures and Tsallis entropy}\\

                            \vspace{2.5cm}

              \normalsize\sf    NIKOS \  \ KALOGEROPOULOS $^\dagger$\\

                            \vspace{0.2cm}
                            
 \normalsize\sf Weill Cornell Medical College in Qatar\\
 Education City,  P.O.  Box 24144\\
 Doha, Qatar\\

                            \end{center}

                            \vspace{2.5cm}

                     \centerline{\normalsize\bf Abstract}
                     
                           \vspace{3mm}
                     
\normalsize\rm\setlength{\baselineskip}{18pt} 

\noindent  We present a path toward determining the statistical origin of the thermodynamic
limit for systems with long-range interactions. We assume throughout that the systems under 
consideration have thermodynamic properties given by the Tsallis entropy. We rely on the 
composition property of the Tsallis entropy for determining effective metrics and measures 
on their configuration/phase spaces. We point out the significance of Muckenhoupt weights, of  
doubling measures and of doubling measure-induced metric deformations  of the metric. 
We comment on the volume deformations induced by the Tsallis entropy composition and 
on the significance of functional spaces for these constructions.      

                             \vfill

                             
                             \vfill

\noindent\rule{8cm}{0.2mm}\\
   \noindent \small\rm $^\dagger$  E-mail: \ \  \small\rm nik2011@qatar-med.cornell.edu\\

\end{titlepage}


                                                                                 \newpage

                                          \normalsize\rm\setlength{\baselineskip}{18pt}

                                                 \centerline{\large\sc 1.   \ Introduction}

                                                                            \vspace{5mm}

The statistical mechanics and thermodynamics of systems involving long-range interactions [1]-[9] has attracted considerable interest 
recently. Such systems were once thought to be ``odd", or ``anomalous" and, as such, outside the realm of applicability or interest of 
conventional thermostatistics. During the last two decades though, they  have been brought at the fore-front for variety of reasons, not the 
least of which is the rich and occasionally unexpected behaviour they exhibit [1]-[9]. The appearance of quasi-stationary states is an 
example of such a behaviour. Moreover, the in-equivalence of the classical ensembles for long-range interactions has provided a 
substantial impetus for the recent re-examination of the foundations of Statistical Mechanics [9], [10]. \\

This note, unlike most work on the topic of long-range interactions, explores the implications of the composition of the 
Harvda-Charvat [11], Dar\'{o}czy [12], Cressie-Read [13], Tsallis [14], [15]  entropy functional, that has also attracted considerable attention 
during the last 25 years.  This entropic form, henceforth called just ``Tsallis entropy", conjecturally describes among 
many other things, the thermodynamic behaviour of systems with long-range interactions. It should be noticed, that despite some 
numerical evidence, such a claim is still conjectural and largely unsettled. Moreover, it should be noticed that a 
considerable amount of work on the statistical and thermodynamic properties of systems with long-range interactions is based on 
the more conventional Boltzmann/Gibbs/Shannon (BGS), rather than on the Tsallis entropy [1]-[7].\\

In this note we analyse the behaviour of systems with long-range interactions from the viewpoint of the Tsallis entropy. We focus, in 
particular, on some formal implications that the Tsallis entropy induced generalised additivity has for such systems [16]-[19]. 
We stress, in particular, the origin and significance of Muckenhoupt weights, of the doubling measures [20],[21] and measure-induced  
metric and volume deformations [22]. 
We take a somewhat ``coarse" view throughout this work: we focus on aspects of these systems that change in a ``uniform" 
manner under the stated deformations. 
By taking this coarse view, we attempt to find common features of systems with long-range interactions viewed from the standpoint  of  
Tsallis entropy.  
Although the treatment in this work has been largely motivated by applications of the geometry of negatively  curved Riemannian 
manifolds [23], it extends to far more general metric-measure spaces which can be fractal [20]-[22] rather than locally Euclidean. \\  
  
Section 2 presents the Ka\v{c} prescription for many-particle systems with long-range interactions whose generalisation leads to the use of
 Muckenhoupt weights. Some of their properties of possible interest are mentioned here. Section 3 relies on a hyperbolic Riemannian 
 metric introduced in our past work and explores  the consequences of the existence of such effective metric and measure structures 
 in configuration/phase space and their deformations. In conjunction with Muckenhoupt weights we are then lead to 
 consider doubling measures on such spaces. Connections with the Tsallis entropy composition are pointed out. Section 4 expands on  
 topics covered in the previous Sections that are loosely connected to each other. Section 5 presents an outlook and topics for current and 
 future work.\\ 
 
                                                                         \vspace{5mm}
              

  \centerline{\large\sc 2. \ Long-range interactions and Muckenhoupt weights}

                                                                            \vspace{5mm}

\noindent{\large\bf 2.1} \ \ The definition of what constitutes a long-range interaction for the purposes of statistical mechanics is essentially 
a criterion of integrability. In the simplest context, consider \  $N$ \  particles of equal mass in \ $\mathbb{R}^n$ \ with Hamiltonian
\begin{equation}
     \mathcal{H} \ = \ \sum_{i=1}^N \ \frac{\vec{p}_i ^2}{2m}  + \sum_{i\neq j} \Phi (|\vec{r}_i - \vec{r}_j|) 
\end{equation}  
Motivated by models of the inter-atomic interactions, such as the Lennard-Jones potential, we assume that the potential energy $\Phi$
has a hard-core repulsive term for inter-particle distances of order 1, and it is asymptotically attractive with behavior 
\begin{equation}
      \Phi (|\vec{r}_i - \vec{r}_j|) \  \sim \  - \frac{C}{|\vec{r}_i - \vec{r}_j|^d}, \hspace{8mm}    |\vec{r}_i - \vec{r}_j| \geq 1 
\end{equation}
where \ $C>0$ \ is a constant. Taking into account the assumed isotropic nature of \ $\Phi (|\vec{r}_i - \vec{r}_j|)$, \ the potential energy per 
particle is approximately given by
\begin{equation} 
     \frac{\Phi}{N} \ \sim \ - C \int_1^\infty \  r^{-d} \ r^{n-1} \ dr
\end{equation}
This quantity converges for \ $d > n$ \ (short-range interactions) and diverges for \ $d < n$ \ (long-range interactions). 
The divergence for \ $d < n$ \ is due to the infra-red behaviour of 
$\Phi$ (2). This is clearly a mean-field approximation like the ones usually employed, ab initio, in atomic, solid-state physics etc. \\

We infra-red regularise (3) by using a cut-off \  $\Lambda$, \  so we get
\begin{equation} 
        \frac{\Phi}{N} \ \sim \ - C \int_1^\Lambda \  r^{-d} \ r^{n-1} \ dr \ = \ - C \ \frac{\Lambda^{n-d} - 1}{n-d} \equiv \Lambda^\ast
\end{equation}
Ref. [15] uses \ $\Lambda = N^\frac{1}{n}$ \ based on the assumption that the \ $N$ \  particles are uniformly distributed within a sphere 
of volume \ $\Lambda$. \  The result of (4) is subsequently used to define reduced quantities such as \ $T^\ast = T/ \Lambda^\ast$ \  
which are in turn used in formulating the equation of state, instead of using the corresponding unreduced quantities which are 
meaningless for long-range interactions. The definition and use of such reduced quantities comes under the label of
 ``Ka\v{c} prescription" [24].\\

A finer distinction arises in thermodynamics [15]. All thermodynamic variables are classified in three,rather than two, broad categories:
the ones coming from energy-type variables, such as the internal, Gibbs and Helmholtz free energies, give rise to pseudo-extensive 
quantities, which scale with \  $NN^\ast$. \  Their Legendre conjugate are pseudo-intensive quantities, such as the temperature, 
pressure etc. Then there are the ``genuinely" extensive ones, which scale with \ $N$ \ such as the volume, mass, entropy etc [15].\\

Another kind of a fine distinction arises in statistical mechanics [6] where one can classify all long-range interactions into two sub-
classes, 
generic systems for which $d>0$ termed strong long-range interacting systems and systems for which \ $d = n + \epsilon$ \ for small 
$\epsilon >0$ termed weak long-range interacting. There are indications  that such systems also exhibit substantially different 
thermodynamic behaviour to warrant separate treatment [6], something that we will not however pursue further in the present work.    \\


\noindent{\large\bf 2.2} \ \ A formal generalisation of (4) can be obtained if we interpret it as an integration of the characteristic function 
 $\chi_{B_\Lambda (0) \backslash B_1(0)}$ of a spherical annulus of inner radius  $1$  and outer radius  $\Lambda$. \ Here \
 $B_r (x)$ \ stands for an open ball centred at $x$ of radius $r$. Then (4) 
 is re-expressed in terms of the characteristic function \  $\chi$ \ in the radial direction as    
\begin{equation}
      \frac{\Phi}{N} \ \sim \ - C  \int_0^\infty r^{-d} \ r^{n-1} \ \chi_{[1, \Lambda]} \ dr 
\end{equation}
where, in general, if \ $U\subset A$ \ then \ $\chi_U = 1$ \ and \ $\chi_{A\backslash U} = 0$. \  In a attempt to make the analysis more 
robust rather than case-dependent, the obvious generalisation of (5) is to 
consider some mollifier/regulator function \ $w: M\rightarrow \mathbb{R}$ \ instead of just the characteristic function \ $\chi$ \ of a set.  
After all, (3)-(5) 
are just rough estimates of the internal energy per particle of the system. So, in essence, one can largely reduce the problem of 
determining the  thermodynamic limit of particle systems with long-range interactions to one of determining if not the exact, at least 
some general approximate features of the desirable functions \ $w$ \ given by    
\begin{equation}
      \frac{\Phi}{N} \ = \ \int_M \Phi (x)  \ w(x) \ dvol_M
\end{equation}
so that (5) remains finite and \ $M = \mathbb{R}^{3nN}$ \ being the configuration space of the system for the rest of this work, 
unless otherwise explicitly indicated.  Such a projection from the phase space is 
possible because the Hamiltonian (1) splits into kinetic and potential terms, the former of which do not involve explicitly the positions 
and the latter of which do not involve explicitly the canonical momenta.   \\

One way to proceed is to consider the ``worst possible scenario" for \ $w(x)$, \
namely the  case in which (6) attains its highest possible value due to \ $w$. \ 
In such a case we obtain the centered Hardy-Littlewood maximal function [20], [25] for \ $w$, \  given by
\begin{equation} 
     \mathcal{M} (w) (\vec{x}) \ = \ \sup_{R>0} \ \frac{1}{vol \ B_R(\vec{x})} \ \int_{|\vec{y}|<R} |w (\vec{x} - \vec{y})| \ dvol_M (\vec{y})  
\end{equation}
We would like such a maximum of the average of \ $w$ \ not to be too much bigger than \ $w$ \ itself  which, after all, originated as a 
regulating function. This has also the desired effect that the value of the internal energy per 
particle \ $\Phi/N$ \ is not too sensitive to the particular choice of \ $w$. \ We enforce this condition by  demanding that 
\begin{equation}      
        \mathcal{M}(w)(\vec{x}) \ \leq \ C' w(\vec{x}), \ \ \ \ \mathrm{for \  almost \ all} \  \  \vec{x}\in M
\end{equation}
Since we are ultimately dealing with integrals of the potential \ $\Phi$ \ in the calculations of thermodynamic quantities, it may be prudent to 
pre-emptively weaken (8) by demanding to hold for  ``almost all" instead of ``all" points \ $\vec{x} \in M$. \ This way we weaken the 
assumptions, thus enlarging the admissible  class of \ $w$, \  as we can exclude ``few" points \ $\vec{x}$ \ that contribute negligibly to such 
integrals. Functions \ $w$ \  satisffying (8) are quite well-known in harmonic (Fourier) analysis [20], [25] and they are called (Muckenhoupt) \ 
$A_1$ \  weights.  It may be worth pointing out that in the definition of the  
maximal function (7), the domains are cubes instead of balls, but such difference is ultimately inconsequential [25] for a ``coarse" 
description of the system. \\

A fundamental corollary of the definitions of the \ $A_1$ \ weights [20], [25] is that such functions \ $w$ \ obey the weak $(1,1)$-type 
inequality
\begin{equation}
         w ( \{ \vec{x} \in M: \ \mathcal{M}(f)(\vec{x}) > \lambda \} )  \ \leq \ \frac{C'}{\lambda} \int_M \ |f(\vec{x})| \ w(\vec{x}) \  dvol_M
\end{equation}
where \ $C'$ \ is the same constant in (8) and (9). To better understand the need for such a weak inequality, as opposed to more familiar 
ones like the ones in the following paragraphs, it may be worth looking into an example. Consider Dirac delta function \ $\delta(x)$ \ at the 
origin of \ $\mathbb{R}$ \ which is equipped with its Lebesgue measure. Then \ $\mathcal{M}(\delta)(x) = \frac{1}{|x|}$ \ which is 
neither integrable  at the origin nor at infinity. Hence the map \ $\delta \mapsto \mathcal{M}(\delta)$ \  is unbounded, so it cannot map 
\ $L^1$ \ to itself via a strong (1,1) inequality [22]. This same conclusion regarding the weak inequality (9) can also be reached for the 
delta function in \ $\mathbb{R}^n$. \\


\noindent{\large\bf 2.3} \ \  We digress from the line of reasoning in this paragraph in order to put some of  above statements in a wider context.  
Let \ $L^p (M)$ \ denote the Banach space of $p$-integrable functions on \ $M$, \  with \ $1\leq p < \infty$ \ namely
\begin{displaymath}
       L^p (M) \ = \ \left\{ f: M\rightarrow \mathbb{R} \ \ \mathrm{such \ that} \ \  \| f \|_p \equiv \left( \int_M |f(x)|^p \ dvol_M \right) ^\frac{1}{p} \ < \ 
       \infty \right\}
\end{displaymath}
This  can be extended to the \ $p=\infty$ \ case by defining
\begin{displaymath}
      L^\infty (M) \ = \ \{ f: M\rightarrow \mathbb{R} \ \ \mathrm{such \ that} \ \ |f(x)| < C, \ \ \mathrm{for \ almost \ all} \ \ x\in M \} 
\end{displaymath}
where \ $C>0$ \  is a constant. The above definitions can be trivially modified to also accommodate complex-valued rather than real-valued 
functions as well as other measures instead of just  \ $dvol_M$.
Assume in the sequel, for simplicity, that \ $M$ \ is a Cartesian power of \ $\mathbb{R}$ \ as, for example, in (6). 
In quantum physics we are interested in determining wave-functions which are elements of \ $L^2 (M)$, for example. \\
 
In the analysis of higher order linear or non-linear microscopic and mesoscopic equations such as the non-linear diffusion 
equation, the non-linear Schr\"{o}dinger equation etc which have attracted considerable interest in the context of Tsallis entropy 
recently [26]-[33], we are dealing with functional spaces that may be dense in or otherwise suitably approximated by \ $L^p(M)$ \ for 
some value of \ $p>1$. \ Following the steps leading to (8) and (9), we reach the conclusion that we are actually interested in 
functions \ $w$ \ which obey the \ $(p,p)$ \ strong integral  inequality
\begin{equation} 
     \int_M \ [\mathcal{M}(f)(x)]^p \ w(x) \ dvol_M \ \leq \ C_p \int_M \ |f(x)|^p \ w(x) \ dvol_M
\end{equation}  
which is the \ $p>1$ \ counterpart of (9). Then the analogue of (8) which is a characterisation of the \ $A_p$ \ (Muckenhoupt) weights \ $w$ \ 
turns out to be [20], [25]
\begin{equation}
      \sup_{B_r} \ \left( \frac{1}{vol \ B_r} \ \int_{B_r} w(x) \ dvol_M \right) \left( \frac{1}{vol \ B_r} \ \int_{B_r} \ [w(x)]^\frac{1}{1-p} 
                                \ dvol_M \right)^{p-1} \ \leq \ C_p \ < \  \infty
\end{equation}
It may be worth noticing that the analogue of (10) does not hold for \ $p=1$ \ and that (9) is the best that someone can do in lieu of (10). 
The relation between \ $A_p$ \ weights for \ $p=1$ \ and \ $p>1$ \ is provided by the realization that if \ $w_1$ \ and \ $w_2$ \ are \ $A_1$ \ 
weights then 
\begin{equation}
       w \ = \ w_1w_2^{1-p} 
\end{equation}
is an $A_p$ weight. Surprisingly, the converse is also true, giving rise to the factorisation of \ $A_p$ \ weights in terms of 
\ $A_1$ \ weights given by (12). We will use properties of the \ $A_p$ \  weights [20], [25] in the sequel as needed.\\


\noindent{\large\bf 2.4} \ \ Calculations involving interacting systems of particles use potentials having the functional form (2), the Lennard-Jones potential being 
an important, rather typical, example. One particular  viewpoint is to combine (2) with the mollifier (regulating) 
function \ $w$ \ and demand the new function to be an \ $A_p$ \ weight, following the line of reasoning that leads to such a conclusion
for \ $w$. \  In effect, we wish to determine for which values of \ $d$, \ $\Phi$ \ in (2) is an \ $A_p$ \ weight. The answer is a straightforward
outcome of simple integrations [25], so rather than actually deriving it, we just quote the final result. Consider, for simplicity, the 
configuration space to be \ $\mathbb{R}^n$. \ It turns out that, for  \ $p>1$, \ $\Phi$ is an \ $A_p$ \ weight if \ $-n < -d < n(p-1)$. For \ $p=1$ \  
which is the case of main interest in the present work, \ $\Phi$ \  turns out to be 
an \ $A_1$ \  weight, if \ $d < n$. \ Since we assume \ $d>0$, \ we see that for all \ $p\geq 1$, \ $\Phi$ is an \ $A_p$ \ weight if and only if \  
$d<n$ \  which is exactly the definition of long-range interactions. To conclude this section we have found that, under our assumptions, 
long range-interactions can be seen as \ $A_p$ \ weights in the configuration space. The viewpoint that combines \ $w$ \ with \ 
$dvol_M$ \ in (6), and its consequences, will be examined in the next Sections.\\

                                                                               \vspace{5mm}
                                                                           

                           \centerline{\large\sc 3. \ Tsallis entropy composition and doubling measures}

                                                                            \vspace{5mm}

\noindent{\large\bf 3.1} \ \ The Tsallis entropy, for a system described by a discrete set of probability of outcomes \ $\{ p_i\}, \  i\in I$, \ where 
\ $I$ \ is an index set whose elements label the probabilities, is given by 
\begin{equation}
        S_q [\{ p_i\} ]  =  k_B \frac{1}{q-1} \left( 1 - \sum_{i\in I} p_i^q \right) \nonumber
\end{equation}
The non-extensive parameter \ $q$ \ can, generically, take any values in \ $\mathbb{R}$.\\

The generalisation of the above expression to the continuous case appeared to be straightforward ([15] and references therein). However it has 
been  recently argued [34] that this may not be so, to the extent of even rendering the concept of Tsallis entropy for continuous systems invalid.
A corollary of the arguments of  [34] is that the application of Tsallis entropy to ``continuous" Hamiltonian systems altogether is non-feasible 
largely for reasons pertaining to the behaviour of the resulting thermodynamic potentials. As a result of [34], some controversy has arisen 
[35]-[37]. Some skepticism is also expressed in [38], [39] which provide some evidence that the Tsallis entropy may be  valid only for 
the restricted range of the non-extensive parameter \ $q\in [0, 1]$. \ Some of the underlying ideas on which [34] relies,  are also examined 
in the more recent [40]. It is also worth mentioning the related work [41]  where the Tsallis entropy is used for continuous systems,
but where some of the difficulties mentioned in the previous works are pointed out and partly addressed. Since the issue of the validity of 
Tsallis entropy for continuous systems raised in [34] does not seem to be fully resolved yet, in our opinion at least, we will use the naive 
extension of the Tsallis entropy shown in the sequel, being fully aware of its potential shortcomings and possible boundaries of its validity. 
Since we will be assuming that $q\in [0,1]$ in the sequel, our conclusions do not explicitly contradict those of [38], [39].\\      
 
The above paragraph's content non-withstanding, assume  a system having as configuration space a Riemannian manifold 
\ $(M, {\bf g})$ \ equipped with a probability measure  \ $d\mu$ \ which is  absolutely continuous with respect to the volume 
element \ $dvol_M$ \ of \  $M$ \ with Radon-Nikodym derivative \ $\rho$. \ Then its Tsallis entropy is  given by  
\begin{equation}      
   S_q [\rho]  = k_B \frac{1}{q-1} \left\{1 - \int_{\Omega} [\rho(x)]^q \ dvol_M \right\}
\end{equation}
For \ $q\rightarrow 1$ \ one recovers the BGS entropy
\begin{equation}   
    S_{BGS} [\rho] = - k_B \int_{\Omega} \rho(x) \log \rho(x) \ dvol_M 
\end{equation}
We will set henceforth \ $k_B = 1$ \ for brevity. 
 Conventionally, two subsystems \ $A, B \subset \Omega$ \ are considered independent if 
\begin{equation} 
     p_{A \cup B} = p_A p_B
\end{equation}
For such subsystems the BGS entropy is additive, namely 
\begin{equation}
      S_{BGS} (A\cup B) = S_{BGS} (A) + S_{BGS} (B)
\end{equation}
 where \ $A\cup B$ \ indicates the system resulting from the combination of  \ $A$ \ and \ $B$. \ 
 The Tsallis entropy though is not additive, as it satisfies
 \begin{equation}   
S_q (A\cup B) = S_q (A) + S_q (B) + (1-q) S_q(A) S_q(B) 
 \end{equation}           
For systems described by the Tsallis entropy, additivity is manifestly restored if a generalized addition \ $\oplus_q$ \ is defined by [16], [17] 
\begin{equation}  
        x \oplus_q y = x + y + (1-q)xy     
\end{equation}            
Then 
\begin{equation}
      S_q (A\cup B) = S_q(A) \oplus_q S_q(B)
\end{equation}    
It took sometime before a generalized product, distributive with respect to the addition (18) was discovered [18], [19]. This turned out to be   
\begin{equation}
   x \otimes_q y = \frac{1}{1-q} \cdot  \left\{(2-q)^\frac{ \log[1+(1-q)x]  \log[1+(1-q)y]}{[\log (2-q)]^2} - 1\right\} 
\end{equation}
Putting together the two generalized binary operations (18) and (20), we set up [19] a one-parameter family of deformations of the set of 
reals  denoted by  \ $\mathbb{R}_q$. \ Such an explicit isomorphism, valid in the range \ $q\in [0,1]$ \ was indicated by  
\ $\tau_q: \ \mathbb{R} \rightarrow \mathbb{R}_q$ \ and was given by    
\begin{equation}
       \tau_q (x) =  \frac{(2-q)^x - 1}{1-q} 
\end{equation}
For future reference, we state its inverse \ $\tau_q^{-1}: \mathbb{R}_q \rightarrow \mathbb{R}$ \ which is 
\begin{equation}
    \tau_q^{-1} (x) \ = \  \frac{\log \{ 1+(1-q) x \}}{\log (2-q)}
\end{equation} 
In [23] we compared, side by side, the composition properties of (13) and (14) by constructing a Riemannian metric on \
 $\mathbb{R}^2$ \ whose metric tensor is given, in rectangular coordinates,  by 
 \begin{equation}
      {\bf g}_h = \left(    \begin{array}{ll} 
                                 1 & 0 \\
                                 0 & e^{-2tx}\\ 
                                \end{array}         \right)
      \end{equation}
where \ $t\in\mathbb{R}$ \ is generally a parameter, whose relation to the non-extensive parameter \ $q$ \ of the Tsallis entropy is given by
\begin{equation}
   t = \log (2-q)
\end{equation}
This metric  turned out to have constant negative sectional curvature [23]
\begin{equation}
    k = - \{ \log (2-q) \}^2
\end{equation}


\noindent{\large\bf 3.2} \ \ At this point a basic question that naturally arises is how important or generic is (23). If one assumes that (23) is an 
effective metric on the 
configuration/phase  space \ $M$ \ of the system, as opposed to being an ``emergent" metric at the level of thermodynamics, then someone 
can reach  several conclusions, without great difficulty. One of them [42]-[44] is that the underlying dynamical system must have largest 
Lyapunov exponent equal to zero, or in other words, that the underlying dynamical system is weakly chaotic. Under the same assumption 
plus the homogeneity of the induced measure, we argued in [45] why we should actually use the escort distributions rather than the naive 
probability ones in the calculations of the thermodynamic averages. Despite the apparent successes of such arguments though, there may 
be a lingering suspicion that the way (23) was constructed  may be ``too artificial" or too ``procedure-dependent" and its conclusions, 
although convenient and, largely, in agreement with simulations,  may not really be generic. So, one could ask  
whether the above mentioned  conclusions  are an artifact of using (23), or whether they are more robust. 
Such skepticism may be strengthened by the fact that not every manifold \ $M$ \  admits a constant negative 
sectional curvature metric reflecting the properties of (23). Consider as such an example the 2-sphere. 
A hyperbolization procedure certainly exists for cell complexes [46] 
but it gives rise to metrics having undesirable regularity properties (``singularities") at some subsets. 
The issue of the existence of sufficiently smooth hyperbolic metrics  on manifolds has been recently, partially, addressed in [47].\\     

One can do with much less stringent requirements though, as was already pointed out in [43], [44]. 
In determining the statistical properties of a system, we are not only interested in the particular system having a given Hamiltonian or 
dissipative deterministic evolution, but also in systems that are sufficiently ``close" to it, initially at least. 
This can be quantified by considering systems having almost the same initial conditions as the one 
under study. One is interested in the evolution of such set of nearby systems, a set of systems having a non-zero volume/measure in $M$. 
Therefore what essentially is of greater interest to Statistical Mechanics, is the behaviour of volumes/measures of $M$  rather than of 
metrics. This is the main topic of ergodic theory and hyperbolic dynamics. In Riemannian manifolds such volumes are controlled by a local 
quantity called  Ricci curvature [48]. Lohkamp's h-principle states that there is no topological obstruction to an $n$-manifold admitting 
a negative Ricci curvature metric, for $n\geq 3$. \ This is in sharp contrast to the \ $n=2$ \ case an example of which was invoked  above. 
Even though this may not resolve all objections about the existence and suitability of negatively curved metrics, 
it relaxes them sufficiently so that we can proceed with arguments relying on the effects on the volume/measure 
rather than on the metrics of \ $M$ \ induced by the Tsallis entropy composition (17), (18).\\       


                                                                \newpage

\noindent{\large\bf 3.3} \ \ We see immediately from (18) is that \ $\oplus_q$ \ is an exponential function for \ $|x| \gg 1, \  |y| \gg 1$. \ 
The exact rate of increase of $\oplus_q$ is controlled by \ $q$. \ 
So the pullback of the Euclidean metric on \ $\mathbb{R}_q$ \ is a logarithmic metric on \ $\mathbb{R}$ \ as can be seen from (22). 
This simple observation was used in  [42], [43] in order to explain why in systems that are described by the Tsallis entropy the largest 
Lyapunov exponent should  vanish. This fact can be generalised to metric spaces [21], [22] with strictly negative upper bound on their 
Alexandrov curvature [48].\\

The  logarithmic map  (22)  has non-trivial effects on the induced measure on \ $M$. \ One such 
implication is explored in [45] which, alongside scaling hypotheses for the induced Riemannian measure, justify the use of the escort rather  
than the ordinary probability distributions in the calculations of the  statistical quantities of interest. Other implications of interest follow.\\


\noindent{\large\bf 3.4} \ \ Consider, for a moment, the familiar situation in Statistical Mechanics: we have a microscopic many-body system given 
by  the Hamiltonian (1). Let its full (unreduced, as opposed to \ $M$) phase space be indicated by \ $\tilde{M}$ \ which, like \ $M$ \ above, 
is assumed to be a Cartesian power of \ $\mathbb{R}$. \ Incidentally, it should be mentioned for completeness, that this condition is not as 
restrictive as it may appear. This relies on the Nash embedding theorem according to which any manifold can be isometrically embedded in 
$\mathbb{R}^n$ for sufficiently large $n$. Let the system have total energy $E$. Then the evolution of the system actually 
takes place in a codimension-1 hyper-surface of $\tilde{M}$, \ which we indicate by \ $\Omega_E$. \ Let the volume $dvol_\Omega$  in \ 
$\Omega_E$ \ be the induced one from \ $\tilde{M}$ \ according to the co-area formula [49]. The simplest (Gibbs) ensemble density \ $\rho$ \ 
on \ $\Omega_E$ \ is the micro-canonical ensemble corresponding to \ $\rho$: constant. According to the ergodic hypothesis, as the time  
\ $t\rightarrow\infty$, \ the evolution of the system will pass through every region \ $D \subset \Omega_E$ \ of finite \ $vol_\Omega$ \ for 
almost all initial conditions. Then following the Birkhoff ergodic theorem, the ensemble averages calculated through \ $\rho$ \ and the 
time averages on \ $\Omega_E$ \ are equal for almost all points of \ $\Omega_E$. \ This is a cornerstone in using the BGS entropy 
(14) which describes the thermodynamic behaviour of such systems. 
What happens however when one is not really interested in $t\rightarrow\infty$ ? Or, in other words, if one is not interested in steady states? 
Or if one is interested in steady states but the relaxation of the system to such states takes a very long time, such as when 
it follows a power law, as with quasi-stationary states? 
Or when ergodicity in the phase space evolution of the system does not hold, as is conjecturally happening for systems described 
by the Tsallis entropy [15]?\\

Suppose, for instance, that one wishes to describe the thermodynamic behaviour  of quasi-stationary states [1]-[3], [50]-[57] 
that appear in systems with long-range interactions.
Then the stationarity condition, which quantifies the equilibrium  distributions, assumed in Birkhoff's theorem is not obeyed. 
A good part of ergodic theory is dedicated to finding asymptotic invariants of dynamical systems, motivated by the approach to equilibrium in 
Statistical Mechanics. 
If we wanted to keep close to the main assumptions that may lead to a generalisation  of Birkhoff's theorem for such cases,  
it is far more reasonable to expect to have to use some ensemble distribution with a density \ $\tilde{\rho}$ \ on \ $\Omega_E$ \ which is 
absolutely continuous with respect to the micro-canonical one. \\

Notice that this regularity assumption on measures, 
which seems reasonable at first glance, is actually non-trivial and may turn out to be too restrictive or even outright wrong: 
indeed consider the Sinai-Ruelle-Bowen (SRB) measures [58] which are special cases of Gibbs measures for dissipative 
dynamical systems. The SRB measures are the most compatible with, and play the role of, the phase space 
volume measure \ $dvol_{\tilde{M}}$ \ in the case of  Axiom A systems [58]. It turns out that they have absolutely continuous densities 
with respect to the restriction of the Riemannian volume of \ $\tilde{M}$ \ on the unstable manifolds [58]  only. So the example of the SRB 
measures shows us that we may run into (naively) unexpected complications when dealing with the statistical description of dynamical 
systems.  Moreover, it is not even obvious that \ $\tilde{\rho}$ \ would even be unique, and if not, what would the physical implications be of 
such behaviour. Having such potential pitfalls in mind, if we still assume the absolute 
continuity of \ $\tilde{\rho}$, \ we see that  they should have the rather generic form
\begin{equation}    
      \tilde{\rho}(x) \ = \ \phi (x) \ dvol_\Omega
\end{equation}
where \ $\phi: \Omega_E \rightarrow \mathbb{R}_+$ \ is a positive deformation function of the micro-canonical density 
of \ $\Omega_E$. \\


\noindent{\large\bf 3.5} \ \ Suppose that we want to get more specific about the possible form of \ $\phi$. \ For this reason, let's assume that the 
system is described by the Tsallis entropy and that (18) is reflected on the metric and measure properties of \ $M$ \ essentially through (21)
and its inverse (22).  Since \ $M$ \ in general is neither homogeneous nor isotropic, \ $\tau_q^{-1}$ \ does 
not act transitively on $M$ \ or on  its tangent bundle \ $TM$. \ But even if \ $M$ \ were homogenous or 
isotropic, one could still reach a similar conclusion, as in the previous case, by assuming that \ $M$ \ has Ricci curvature 
uniformly bounded from below and use the resulting concentration of measure argument for its volume [59]. 
Alternatively, for compact \ $M$, \ one could check whether the first nontrivial eigenvalue of the Laplacian of \ $M$ \ would 
approach infinity in the thermodynamic limit \ $N\rightarrow\infty$ \ which would signify the concentration of the volume 
sequence as a Levy family [59]. As a conclusion then, we should be aware that the measure properties of \ $\tilde{\rho}$ \ 
may not be very well encoded by the (Hausdorff) dimension of \ $M$ \ but other asymptotic invariants may be used to that end.\\ 

In general, the dilatation of a map $f: X\rightarrow Y$ between metric spaces (not necessarily Riemannian manifolds) with respective 
distance functions \ $d_X, \ d_Y$ \ is defined by [21]
\begin{equation}
    dil(f) \ = \ \sup_{x\neq y} \ \frac{d_Y (f(x), f(y))}{d_X(x,y)}, \hspace{8mm} \forall \ x,y \in X
\end{equation}
with the local dilatation at \ $x\in X$ \ being 
\begin{equation}
   dil_x(f) \ = \ \lim_{r \rightarrow 0} \ dil (f\big|_{B_r(x)})
\end{equation}
The dilatation quantifies the relative distortion of distances by \ $f$. \ In our case \ $f = \tau_q$ \ and \ $X = Y = M =\mathbb{R}^{n'}$ \
for some \ $n'\in\mathbb{N}$. \ Then  
\begin{equation}
        dil (\tau_q) \ = \ \frac{|\tau_q(x) \ominus_q \tau_q(y)|}{|x-y|}
\end{equation}
where \ $|\cdot |$ \ stands for the Euclidean distance in \ $\mathbb{R}^{n'}$. \  Since \ $\tau_q$ \ is an isomorphism, or by a direct calculation,
\begin{equation}
   |\tau_q (x) \ominus_q \tau_q(y)| \ = \ \frac{(2-q)^{|x-y|} - 1}{1-q} 
\end{equation}
the local dilatation at \ $x \in M$ \ becomes 
\begin{equation}
     dil_x(\tau_q) \ = \ \frac{\log(2-q)}{1-q}
\end{equation}
Therefore, by choosing 
\begin{equation}
    \phi (x) \ = \ \{dil_x(\tau_q)\}^\alpha
\end{equation}
we do not gain anything as such an identification amounts to a re-scaling of the volume element of \ $M$ \ by a global constant.
Here \ $\alpha$ \ is an effective dimension of \ $\tilde{\rho}$ \ and generally  \ $\alpha \neq dim M$ \ for the reasons indicated in the 
previous paragraph. Therefore another approach has to be followed in narrowing down the form of \ $\phi$.\\ 

To make some progress in determining properties of $\phi$ we are motivated by asking what is the difference between Euclidean and 
hyperbolic (negative sectional curvature) volume elements. In the former case, the volume of the ball \ $B_R(x)$ \  in \ $\mathbb{R}^n$ \ 
is given by
\begin{equation}
    vol \ B_R (x) \ \sim \  \mathrm{const}_n \ R^n
\end{equation}
In the latter case the volume of a ball in a hyperbolic space is 
\begin{equation}
    vol \ B_R(x) \ \sim \  \mathrm{const}'_n \ e^R, \hspace{5mm} R\rightarrow\infty
\end{equation}
where the constants depend on the (Hausdorff) dimension $n$ of each space. Naturally, both (33) and (34) give the same result if 
\ $R\rightarrow 0$ \ as all manifolds are locally isometric to Euclidean spaces. 
The differences in the volume expression between the Euclidean 
and hyperbolic spaces become more dramatic, and of interest to us, as \ $R\rightarrow\infty$, \ as can be seen by comparing (33) and (34). 
It should be noted that the result of (34) also holds for spaces of constant negative 
curvature \ $k<0$ \ by substituting \ $-k R$ \ instead of \ $R$ \ as argument of the exponential in (34).   
We see then that (22) maps (34) into (33). An implication of this, that has been known for a while [60] is that the Tsallis entropy describes 
systems whose phase space volume growth is polynomial rather than exponential [60]-[62]. 
As result, and in order to pinpoint non-trivial implications of (21), (22), and by extension of (18), it would be desirable  
to find a basic property of \ $M$ \  that is satisfied by one of (33), (34) but not by the other.\\


 \noindent{\large\bf 3.6} \ \ Motivated by the statement [20], [25] that for any \ $A_p$ \ weight \ $w$ \ in $\mathbb{R}^n$, \ the deformation of \ 
 $dvol_M$ \ given by 
 \ $w(x) dvol_M$ \ is a doubling measure, we choose the sought after property to be the ``doubling property" [20] - [22]. 
 This  can be formulated for a general metric space $M$ and not just for $\mathbb{R}^n$, so we give this more general definition here. 
 A metric space is called doubling if there is a constant \ $c>1$ \ such that every set of diameter 
\ $2R$\  in this space can be covered by at most \ $c$ \ sets of diameters at most \ $R$. \ 
This can be thought as some kind of finite-dimensionality condition, following the definition  of topological dimension [22]. Alternatively, 
it can be interpreted as stating that all snapshots of the metric space are bounded [22]. This doubling condition is clearly true for the 
Euclidean \ $\mathbb{R}^n$ \ but not for the hyperbolic space. The reason is simple: it takes exponentially many balls to cover a ball whose 
center is close to the ideal boundary of the hyperbolic space. More intuitively, there is far ``too much space" close to the ``infinity" of a 
hyperbolic space to be controlled in a polynomial manner as is demanded by the definition of a doubling space. 
The ``doubling" concept can be extended to 
measures \ $\mu$ \ in a metric-measure space as follows: \ $\mu$ \ is a doubling measure if there is a constant $c'>0$ such that  
\begin{equation}     
    \mu (B_{2r}(x)) \ \leq \ c' \mu (B_r(x))
\end{equation}
The relation between these two definitions is (almost) what one may intuitively expect: a space with a doubling measure is doubling. 
Almost conversely, every complete doubling space carries a doubling measure [21]. 
Notice however that there are doubling spaces that do not carry doubling measures [63]. \\


\noindent{\large\bf 3.7} \ \ Some implications of these concepts that may make them somewhat more interesting for (Statistical) Physics: 
one can ask under what 
conditions is the inter-atomic potential (2), seen as measure on \ $\mathbb{R}^{n'}$, \ doubling. Assuming that \ $d>0$, \ we can 
straightforwardly check that the doubling condition amounts to the condition \ $d < n$. \ This is exactly the same as the condition 
for \ $\Phi$ \ to give rise to an \ $A_p$ \ weight as was pointed out in the last paragraph  of Section 2. 
Hence we have two characterizations, which are related but not identical, for long-range interactions: as $A_p$ weights 
and as doubling measures on \  $\mathbb{R}^{n'}$. The doubling condition also restricts the form that \ $\phi$ \ in (26) can have: the 
requirement for \ $\phi$ \ is to be such that \ $\tilde{\rho}(x)$ \ is doubling. The doubling condition is a uniformity condition on measures, 
playing a role for measures analogous to that of re-scalings for metrics. Many fractal spaces possess doubling measures. For this 
reason and due to the relation between the foundations of the Tsallis entropy with fractals [14], [15]  doubling measures may prove useful in
further exploring the relation between Tsallis entropy and long-range interacting systems.\\


 \noindent{\large\bf 3.8} \ \ The above line of reasoning may seem to indicate that long-range interactions  would imply a power-law phase space volume 
 growth rate which should be associated, or demand, the use of Tsallis entropy in their statistical description. 
However, one should be careful in making such extrapolations. There are several non-trivial conditions that are assumed in the 
above arguments to be valid: we would be quite surprised if all such assumptions were valid in the evolution of even the simplest 
models of long-range interactions throughout their full parameter space. To what extent these assumption are correct or where they may 
fail can only be decided by analyzing specific models. What can be stated without much doubt though is that (18) carries with it traces of 
origins of the Tsallis entropy in the (multi-)fractal formalism, as many fractals are carrying doubling measures [21], [22].    \\       


\noindent{\large\bf 3.9} \ \ A question that is motivated by the situation of the SRB measures is on whether such doubling measures are flexible enough of our 
purposes? It is possible, for instance, a doubling measure even on \ $\mathbb{R}^{n'}$ \ to be singular with respect to volume, 
mirroring the behaviour of the SRB measures on the stable sub-manifolds of phase space? The answer to this is affirmative 
and is provided via Riesz products, which are some sort of multiplicative Fourier series. Beyond this existence statement, we will not expand 
on this rather technical topic as it may not be of interest in physics, and we will refer to [20], [25]  for further information and references.\\


\noindent{\large\bf 3.10} \ \ There are three more reasons for considering doubling measures in spaces related to a metric 
structure induced from (18). 
First, it is possible to carry over familiar concepts from  Fourier (harmonic) analysis on such spaces [20]-[22], [25]. This is invaluable 
for numerous aspects of classical and quantum physics. Second, there is a theorem [64] stating that  a metric space \ $(M, d)$ \ is 
doubling if and only if its ``snowflake space" \ $(M, d^\epsilon ), \ \ 0< \epsilon < 1$ \ admits a bi-Lipschitz embedding into some 
Euclidean space. As a reminder, an embedding \ $f: M_1 \rightarrow M_2$ \ between metric spaces \ 
$(M_1, d_1)$  and  $(M_2, d_2)$  is $C-$ bi-Lipschitz, for $C>1$,  if 
\begin{displaymath}   
     \frac{1}{C} \ d_1 (x,y) \ \leq \ d_2 (f(x), f(y)) \ \leq \ C d_1 (x,y), \hspace{8mm} \forall \ x, y \in M_1   
\end{displaymath}
Bi-Lipschitz equivalence of two metric spaces is one of the several available statements about the ``coarse equivalence" of such spaces.  
Bi-Lipschitz embeddings distort distances but in  a uniformly controllable way. Such maps are not too ``exotic" or ``pathological"  either: 
indeed, they are almost everywhere differentiable [49]. The ``snowflake transformation" alluded to above is a standard device in the analysis 
and geometry of metric spaces which is used, among many other thing for providing (counter-) examples for several conjectures
[21], [22]. In the present context Assouad's theorem given above provides a characterisation of doubling spaces when compared to 
Euclidean spaces from the viewpoint of coarse geometry. Someone could ascribe to such statements, in part, the fact that it is posible to use 
some common concepts and approaches for analyzing both fractal and Euclidean spaces, or in short, for exploring the consequences of  
(18) versus the usual addition. \\

Third, consider a sequence of metric spaces that are all doubling with the same doubling constant. 
Moreover, assume that all such spaces are proper, namely that all their closed balls are compact.
Such a sequence has a convergent subsequence, where convergence is meant in the Gromov-Hausdorff (GH) sense [21], [65]. 
This can be thought  as a closure and stability result since the limit of a subsequence belongs to the same class of 
spaces, having the same features, as the elements of the sequence. The definition and properties of the Gromov-Hausdorff 
metric would take us too far afield and we refer to  [65] for an excellent treatment. It is sufficient to just comment that it captures 
the idea of distance if each (compact) set can be uniformly approximated by a countable set of points (an $\epsilon$-net). 
Therefore the GH convergence provides a ``coarse" way for sequences of spaces to converge. Such convergence is of substantial 
interest when one has to consider the thermodynamic limit, for instance. Such a sequence of geometric structures 
can be produced by the terms of a perturbative expansion of the entropy/free energy functionals. It is quite reassuring, technically,
 to know that in a system without any phase transitions the limit shares the properties of the elements of the sequence that gives rise to it. \\  

 
\noindent{\large\bf 3.11} \ \ In the above, we discussed implications of  (18) on the measure/volume density $\tilde{\rho}$ of phase space $M$ 
without explicitly perturbing the underlying metric. A question that arises  is whether such volume/measure deformations can be 
accounted for through variations of the underlying metric. In some sense this is a stability problem: what kind of metric 
deformations, if possible at all, can account for such measure changes? Given a metric with distance function $d$ 
(in the cases above it is the Euclidean one, but the construction is far more general) and a doubling measure \ $\mu$ \ on some space, 
one can deform the metric using $\mu$ by constructing a deformed metric [22] whose distance function is given by    
\begin{equation}
    D(x,y) \ = \ \{ \mu (\bar{B}_{d(x,y)} (x)) + \mu (\bar{B}_{d(x,y)} (y)) \}^\beta
\end{equation}
where the overbar denotes a closed ball and \ $\beta > 0$. \ This deformation is rather benign in \ $\mathbb{R}^{n'}$ \ in that if the 
metric space is \ $\mathbb{R}^{n'}$ \ and \ $\beta = 1/n'$ \ then one gets back the Euclidean metric. In general however, 
\ $D(x,y)$ \ is not even a metric, as it satisfies 
\begin{equation}
      D(x,z) \ \leq \ k \ (D(x,y) + D(y,z))
\end{equation}
for some \ $k>0$ \ and \ $x,y,z$ \ any three points of the space, instead of the triangle inequality. However, it was shown in [66], that 
there is a metric \ $\tilde{d}$ \ and constants \ $\tilde{c} >0$ \ and \ $s\geq 1$ \ such that 
\begin{equation}  
       \frac{1}{\tilde{c}} \  \{ \tilde{d}(x,y) \}^s   \   \leq   \ D(x,y)   \    \leq    \     \tilde{c} \ \{ \tilde{d}(x,y) \}^s   
\end{equation}
so, roughly speaking, \  $\tilde{d}$ \ and \ $D$ \  do not give substantially different results. Therefore, from a rough/coarse viewpoint, 
the deformations of the metrics due to doubling measures are rather benign. This does not mean, of course, that they are not 
substantial or that have no appreciable physical implications. But we can see as the role of such deformations as pointing out 
where to search for more detailed information: since in the simplest case of \ $\mathbb{R}^{n'}$ \ such changes result in local 
re-scalings of the metric, which are conformal transformations, then one may wish to consider such conformal changes as the effect 
of (18) on phase space, in general. This is clearly compatible with, and supports as an additional line of argument, 
 the induced conformal change of measure expressed in (26).\\ 
 

                                                                  \newpage


     \centerline{\large\sc 4. \  \  Further implications }

                                                                     \vspace{5mm}
  
 In this section we expand on some topics touched upon in the previous paragraphs, which however may be only loosely 
 associated to each other.\\ 
  
\noindent{\large\bf 4.1} \ \ The concept of ergodicity breaking is quite frequently mentioned in the thermodynamics  of systems of 
long-range interactions as a possible explanation for some of their more unusual properties. We would like to point to that     
using \ $A_p, \  \  1\leq p \leq\infty$ \ weights generalises in a specific way some of the points associated to ergodicity.
If one considers such an \ $A_p$ \ weight \ $w(x)$ \ as a deformation of the Euclidean volume in \ $\mathbb{R}^{n'}$, 
\ then it turns out [20], [22], [25] that there are constants \ $0 < \delta_1 < 1$, \  $0<\delta_2$ \  such that any measurable subset
 \ $U \subset B_r(x)$ \ satisfies 
\begin{equation}
    vol \ U \ \leq \ \delta_1 \ vol \ B_r(x) \  \  \implies \  \ w(U) \  \leq \delta_2 \ w(B_r(x))  
\end{equation}
Hence one of the basic ingredients of the Birkhoff ergodic theorem, which enters in the volume/measure average of the 
ensemble distribution is almost preserved: even under \  $A_p$ \  weights a subset roughly maintains the percentage of the overall 
measure of the set that it occupies.  Hence a generalised ergodic Birkhoff theorem in an \ $L^p$ \ sense may make sense: under some 
conditions, yet to be specified the \ $L^p$ \  average over the evolution parameter ``time" may equal the \ $L^p$ \ ensemble (26) average 
\begin{equation}
         \lim_{T\rightarrow\infty} \  \frac{1}{T} \int_0^T  |f(t)|^p \ dt  \ \ = \ \ \int_{\mathbb{R}^{n'}}  |f(x)|^p \ \phi (x) \ dvol        
\end{equation}        
The difficult part  in establishing (40) is finding sufficiently general conditions that guarantee the existence of such an equilibrium distribution 
\ $\tilde{\rho}$ \ as in (26) and also examine its uniqueness under such conditions. 
Such a relation would then directly reflect through the deformation function \ $\phi$ \ 
properties of the Tsallis entropy composition (18). The use of \ $L^p$ \ rather than of \ $L^2$ \ integrality in (40) 
could be ascribed to either the non-linear nature of the underlying mesoscopic description or to the lack of Markovian evolution 
of the system that could warrant the use of fractional derivatives and integrals  [26]-[33]. \\


\noindent{\large\bf 4.2} \ \ The important property of (22) is that it is essentially a logarithmic function. One implication is:
 it is often stated that for long-range interactions the energy of a sub-system is no longer additive, since the boundary 
 contribution of the subsystem cannot be ignored even in the thermodynamic limit. This presents a substantial challenge in obtaining 
 the correct thermodynamic behaviour of the system [1]-[6]. For short-range interactions the bulk contributions to  the internal energy 
 scale as the volume of the system \ $V$, \ whereas the surface contributions scale as the surface area, which for sufficiently smooth  
 surfaces, scales as  \ $V^{\frac{n-1}{n}}$. \ For \ $V\rightarrow\infty$ \ the bulk contributions dominate. Now consider the effect of (21).
 We see that \ $\log V$ \ and \ $\log V^{\frac{n-1}{n}}$ are proportional to each other so they scale the same way in terms of  $V$.   
 Now consider a long-range interaction. Suppose that the internal energy contributions due the surface $\Phi_S$ and the bulk $\Phi_V$ 
 are of comparable order of magnitude but not equal. Assume that  \ $\Phi_V > \Phi_S$. \ Then 
 \begin{equation}  
      \exp \Phi_V \  \gg  \  \exp \Phi_S  
 \end{equation}
 especially in the ``thermodynamic" limit \ $n\rightarrow\infty$. \ So the exponential map (21) can be used to map the effect of a 
 long-range into that of a short-range interaction as far as their bulk and surface contributions are concerned. It may not be unreasonable 
 to expect that  one should be  able to deal with some properties of long-range interactions by applying the BGS entropy on the 
 exponentiated versions of long-range interactions, or by straightforwardly applying the Tsallis entropy-induced thermodynamics [15]  
 to such systems. \\
 
 
\noindent{\large\bf 4.3} \ \  A second implication of the logarithmic form of (22), is the following: such a map has the form 
\begin{equation}
     \log P(x), \hspace{5mm} P(x): \mathrm{polynomial}
\end{equation}
Such logarithmic functions, in general, do not belong to any of the Banach spaces \ $L^p(\mathbb{R}^n)$ \ for two possible reasons: 
their potential sources of non-integrabilitiy can arise either from the zeroes of $P(x)$ or from the unboundedness of the logarithmic function 
``at infinity". They do however belong to a more general functional space, which is that of the functions of bounded mean oscillation ($BMO$)
whose definition is as follows [20], [25]: Let \ $f: M\rightarrow \mathbb{R}^n$ \ be locally integrable and let \ $f_B$ \ indicate the mean value 
\begin{equation}   
       f_B \ = \ \frac{1}{vol \ B_r(x)} \ \int_{B_r(x)} \ f(y) \ dvol_M (y)
\end{equation} 
Then \ $f$ \ belongs to \ $BMO(M)$ \  if 
\begin{equation}
     \| f \|_{BMO} \ \equiv \ \sup_{B_r(x)} \frac{1}{vol \ B_r(x)} \ \int_{B_r(x)} |f(y) - f_B| \ dvol_M(y)  \ <  \ \infty
\end{equation} 
for all balls \ $B_r(x) \subset M$. \ One can immediately check that \ $\log |x|$ \ is in \ $BMO(\mathbb{R}^n)$ \ but not in \
 $L^\infty (\mathbb{R}^n)$, \ for instance. The $BMO(\mathbb{R}^n)$ \ spaces are of particular interest for our purposes for one 
 additional reason: it turns out that \ $w\in A_p(\mathbb{R}^n)$, \ implies that \ $\log w \in BMO(\mathbb{R}^n)$. \ 
 The $BMO$ spaces have many desirable properties such as  invariance under translations, dilatations etc [20], [25]. 
 It is also worth observing that $BMO$ functions $f$ are nearly bounded in \ $\mathbb{R}^n$ \ [20], [25]:
 \begin{equation}
       \int_{\mathbb{R}^n}  \  \frac{| f(x) - f_{B_1(0)} |}{(1+|x|)^{n+1}} \  dvol_{\mathbb{R}^n} \ \leq \ const \ \| f \|_{BMO}
 \end{equation}    
So, in a way, the role of (18) is to turn unbounded integrals of the form (3) into almost bounded integrals. This may be one justification for 
the underlying emergence and use of the Ka\v{c} prescription in regularizing unbounded integrals and in normalising thermodynamic 
potentials  and variables appearing in systems with long-range interactions.   \\
 
The \ $BMO(\mathbb{R}^n)$ \ functions are not ``too different" from the integrable ones. This can be quantified in two ways:
One is to observe that if in (44) the terms of the integrand \ $f_B$ \ is omitted, then we would get  the definition of integrable 
functions of class $L^\infty$ indicated in Section 2.   Another way is to invoke a theorem of John and Nirenberg [20] stating that if 
\  $f\in BMO(\mathbb{R}^n)$ \ then \ $f$ \ is locally integrable in the \ $L^p$ \ sense. Yet a third one appears in the next paragraph 
and is their exponential integrability, some implications of which for statistical field theory are presented in the sequel. \\
 
A different way to look at use of  $BMO$ spaces is to use the exponential integrability of $BMO$ functions 
[25]. In quantum and statistical field theories, we are interested, to a large extent, with the equilibrium properties of systems of many 
(effective) degrees of freedom. Following Gibbs' ideas about the equilibrium description of such systems, we are calculating, in the 
canonical approach, the partition function
\begin{equation}
    \mathcal{Z}(\beta )  \ = \ \int \ [\mathcal{D} \phi ] \ e^{-\beta S [\phi ]}
\end{equation}     
where \ $\phi$ \ stands for the set of all fields in the action/free energy $S$. There  is an ad hoc  choice to be made about the 
measure, usually taken to be Gaussian, \ $[\mathcal{D} \phi]$, \ on the space of fields. To make sense of (46) various 
techniques are employed. One of them is to simplify the system by using discretizations. This approach aims to reduce the 
uncountably infinite set of field configurations in (46) to something more manageable. A way to do this is to replace functionals by  
integrable functions in the exponential of (46). The same approach can be  applied to systems with long-range 
interactions. Assuming  that the essentially BGS-based approach of (46) can be used for such systems, it is  imperative to use in (46) 
functions that are exponentially integrable. The use of \ $BMO$ \ functions guarantees this result. Indeed the John-Nirenberg inequality 
states that [20] if \ $f \in BMO (\mathbb{R}^n)$, \ then there are constants \ $c_1>0, \  c_2>0$ \ such that for all \  $\alpha >0$ \ and for all 
balls \ $B\in\mathbb{R}^n$ \ we have     
\begin{equation} 
     vol  \{ x \in B: \ |f(x) - f_B| > C \}   \  \leq    \     c_1 e^{-\frac{c_2 \alpha }{ \| f \|_{BMO}}}  \ vol_B
\end{equation} 
So, in some sense the \ $BMO(\mathbb{R}^n)$ \ is a natural space from which to draw the functions needed to regularise (46). \\


\noindent{\large\bf 4.4} \ \ Suppose that someone uses a Tsallis entropy-based functional, instead of (46), for a system with long-range 
interactions. This approach has actually has being strongly advocated [15] (and references therein). 
The analogue of the canonical partition function (46) would now  be [15]
\begin{equation} 
         \widetilde{Z}_q (\beta_q) \ = \  \int \ [\widetilde{\mathcal{D}}\phi ] \ e_q^{-\beta_q S[\phi]}
\end{equation}
where the $q$-exponential function \ $e_q: \mathbb{R} \rightarrow \mathbb{R}_+$ \ is, by definition
\begin{equation} 
    e_q(x) \ = \ [1+(1-q)x]_+^\frac{1}{1-q}, \hspace{12mm}  [y]_+  = \max \{y,0 \}
\end{equation}
and \ $\beta_q$ \ should be considered as a Lagrange multiplier / parameter, in lieu of the temperature of a system. 
There are several subtleties related to (48), some of which are discussed in [67]-[69], and some fundamental, 
but largely unresolved, issues [70]-[72]. Due to (3) and (49) the $q$-exponential term in (48) diverges for systems with 
long-range interactions. A possible way around 
this difficulty is to carefully choose the measure \ [$\widetilde{\mathcal{D}} \phi$]. \  It should be noted though that different choices
of this measure will almost certainly result in different values of the physical quantities predicted by the model under study. 
So this approach may not be particularly desirable as it potentially lacks predictability, at least on an approach relying on first principles.\\
  
An alternative way to deal with this difficulty is to use the logarithmic map (22). Since there is no difference, algebraically, between 
\ $\mathbb{R}$ \ and \ $\mathbb{R}_q$ \ due to the isomorphism (21), we can always assume that the model is initially formulated at 
\ $\mathbb{R}_q$. \ We can then use  (22) to pull back the results to \ $\mathbb{R}$, \ perform the thermodynamic analysis in 
\ $\mathbb{R}$ \ and then use (21) to push the results forward to \ $\mathbb{R}_q$ \ on which the model was assumed to have been 
initially formulated. What we gain from this roundabout approach, is that the singularities in the exponent of (46) due to long-range 
interactions become much milder, therefore more manageable, due to the subsequent application of (22).  
Even though one starts with the BGS expression (46), one uses in a very substantial way (22) which is induced by the Tsallis entropy.
This way we map non-integrable functions to functions belonging to \ $BMO(\mathbb{R}^n)$ \ as was also pointed out in Subsection 4.3.
It is not clear, a priori, this ``hybrid" approach will give results more compatible with experimental data than by just using (46) or (48), but it 
may a viable possibility, especially since the direct calculation of (48) seems to be particularly difficult even for the simplest models [73] and 
moreover, its convergence rate to a (quasi-) stationary distribution of interest should be rather slow, due to the power-law form of (48). \\   
     
                                                                          \vspace{5mm}
 

                                        \centerline{\large\sc 5. \  \  Conclusions \ and \ Outlook}

                                                                     \vspace{5mm}
   
 In this work we have explored some analytic consequences of the Tsallis entropy composition (17) for systems with long-range interactions. 
 The relations between Tsallis entropy and long-range temporal and spatial correlations have been frequently speculated in the literature 
 [15]. The evidence for such a connection however has been surprisingly weak, at least by using analytic arguments (versus numerically  
 based data fittings). 
 Our viewpoint has a certain slant toward geometric constructions, following those of [22], that are valid and useful not only 
 in the context of Riemannian configuration/phase spaces but also in more general  measure spaces such as  
 homogeneous and fractal spaces.  \\ 

We saw that Tsallis entropy composition (17)  is rich enough to allow it to use several well-known concepts in 
(harmonic) analysis: Muckenhoupt weights, doubling measures, $L^p$ and $BMO$ spaces, etc. As a result, one may use them to make 
some progress toward a better understanding of features of systems with long-range interations such as the Ka\v{c} prescription etc. 
We have moreover seen that conformal deformations of the (Euclidean) metric or, more generally, of the configuration space volume (26)
should play a prominent role in the analysis of such systems. Such systems seem to also have some features that are quite robust and 
maintain their character under mild deformations of the metric or the volume elements (36), at least in a coarse/bi-Lipschitz sense.
They also point out at conjectural extensions of  ergodic theory in a polynomial rather than an exponential/logarithmic function setting  \\

It would  be desirable if one could map the behavior of long-range systems to those of short-range interacting systems non-trivially.
It is unavoidable that such a map would miss important features of both classes of systems. However it may be an alternative step into 
a better analytical understanding of the statistical mechanics of long-rang interactions by mapping them into the far better understood 
systems with short-range interactions. From the above analysis one may be tempted to point out that employing the 
maps (21), (22) may help make some progress toward this goal, if valid at all.\\

Apart from the above ideas, whose tip has barrely been scratched in the present work, but may warrant further exploration, 
another direction of study would be to better understand the origin, scope and properties of the  deformations of the volume 
the configuration space (26), not only in Riemannian but also in the context of fractal spaces as motivated by the Tsallis entropy. 
This is a topic of a forthcoming work [74].\\   
  
                                                                          \vspace{5mm}
 
 
                                                \centerline{\large\sc Acknowledgement}  

                                                                           \vspace{3mm}

We would like the thank the referee for bringing to our attention some of the subtleties  involved in the
definition of the Tsallis entropy for continuous systems and for pointing out [34].\\   
                                                 
                                                                           \vspace{8mm}

 
                                                                                                                              
                                                        \centerline{\large\sc References}
 
                                                                          \vspace{5mm}
 
 \noindent [1] \  A. Campa, T. Dauxois, S. Ruffo, \ \emph{Phys. Rep.} {\bf 480}, \ 57 \ (2009).\\ 
 \noindent [2] \  A. Campa, A. Giansanti, G. Morigi, F. Sylos-Labini, \ (Eds.), \  \  \emph{Dynamics and \\
                               \hspace*{6mm} thermodynamics of systems with long-range interactions: Theory and Experiments}, \\
                               \hspace*{6mm}  Amer. Inst. Phys. Conf. Proc. \ {\bf 970}, \  (2008).\\ 
 \noindent [3] \ T. Dauxois, S. Ruffo, L.F. Cugliandolo (Eds.), \ \emph{Long-Range Interacting Systems}, \\ 
                                 \hspace*{6mm}  Les Houches Summer School 2008, \ Session XC, \ Oxford Univ. Press, \ Oxford (2009).\\ 
 \noindent [4] \ P.-H. Chavanis, \ \emph{Eur. Phys. J. Plus} {\bf 127}, \ 19 \ (2012).\\
 \noindent [5] \ P.-H. Chavanis, \ \emph{Kinetic theory of spatially homogeneous systems with long-range \\ 
                                      \hspace*{5mm} interactions: II. Basic equations} \ {\sf arXiv:1303.0998}\\
 \noindent [6] \ F. Bouchet, S. Gupta, D. Mukamel, \ \emph{Physica A} {\bf 389}, \ 4389 \ (2010).\\
 \noindent [7] \ T.N. Teles, F.P. da C. Benetti, R. Pakter, Y. Lavin, \ \emph{Phys. Rev. Lett.} {\bf 109}, \ 230601 \\
                                     \hspace*{6mm} (2012).\\
 \noindent [8] \ S. Gupta, D. Mukamel, \ \emph{Quasistationarity in a long-range interacting model of particles\\
                                      \hspace*{6mm} moving on a sphere},   \ {\sf arXiv:1309.0194}\\
 \noindent [9] \ O. Cohen, D. Mukamel, \ \emph{Ensemble inequivalence: Landau theory and the ABC model}, \\
                                      \hspace*{6mm} {\sf arXiv:1210.3788}\\
 \noindent [10] \ R. Chetrite, H. Touchette, \ \emph{Phys. Rev. Lett} {\bf 111}, \ 120601 \ (2013).\\  
 \noindent [11] \ J. Havrda, F. Charvat, \ \emph{Kybernetika} {\bf 3}, \ 30 \ (1967).\\
 \noindent [12] \ Z. Dar\'{o}czy, \ \emph{Inf. Comp. / Inf. Contr.} {\bf 16}, \ 36 \ (1970).\\
 \noindent [13] \ N.A. Cressie, T. Read, \ \emph{J. Roy. Stat. Soc. B} {\bf 46}, \ 440 \ (1984).\\ 
 \noindent [14] \ C. Tsallis, \ \emph{J. Stat. Phys.} {\bf 52}, \  479 \ (1988)\\ 
 \noindent [15] \  C. Tsallis, \ \emph{Introduction to Nonextensive Statistical Mechanics: Approaching  \\
                           \hspace*{8mm}   a Complex World}, \ Springer \ (2009)\\
 \noindent [16] \ L. Nivanen, A. Le Mehaut\'{e}, Q.A. Wang, \ \emph{Rep. Math. Phys.} {\bf 52}, \ 437 \ (2003).\\ 
 \noindent [17] \ E.P. Borges, \ \emph{Physica A}  {\bf 340}, \  95 \  (2004).\\
 \noindent [18]  \ T.C. Petit Lob\~{a}o, P.G.S. Cardoso, S.T.R. Pinho, E.P. Borges, \  \emph{Braz. J. Phys.} {\bf 39},\\
                                  \hspace*{8mm}   402 \  (2009). \\          
 \noindent [19]  \ N. Kalogeropoulos, \ \emph{Physica A} {\bf 391}, \ 1120 \ (2012). \\
 \noindent [20]  \ E.M. Stein, \ \emph{Harmonic Analysis: Real-Variable Methods, Orthogonality and Oscillatory\\
                                   \hspace*{8mm} Integrals}, \  Princeton University Press, \ Princeton, NJ \ (1993).\\ 
 \noindent [21] \ J. Heinonen,  \emph{Lectures on Analysis on Metric Spaces}, Springer-Verlag, New York (2001).\\
 \noindent [22] \ S. Semmes, \ \emph{Some Novel Types of Fractal Geometry}, \ Clarendon Press, \ Oxford (2001).\\
 \noindent [23] \ N. Kalogeropoulos, \  \emph{Physica A} {\bf 391}, \ 3435 \ (2012).\\
 \noindent [24] \ M. Ka\v{c}, G.E. Uhlenbeck, P.C. Hemmer, \ \emph{J. Math. Phys.} {\bf 4}, \ 216 \ (1963).\\
 \noindent [25] \  L. Grafakos, \ \emph{Modern Fourier Analysis, \ 2nd Ed.} \ Springer, New York \ (2009).\\
 \noindent [26] \ L.C. Malacarne, R.S. Mendes, I.T. Pedron, E.K. Lenzi, \ \emph{Phys. Rev. E} {\bf 63}, \ 030101(R) \\
                                   \hspace*{8mm} (2001).\\
 \noindent [27] \ I.T. Pedron, R.S. Mendes, L.C. Malacarne, E.K. Lenzi, \ \emph{Phys. Rev. E} {\bf 65}, \ 041108 \\
                                   \hspace*{8mm} (2002).\\
 \noindent [28] \ E. K. Lenzi, L.C. Malacarne, R.S. Mendes, I.T. Pedron, \ \emph{Physica A} {\bf 319}, \ 245 \ (2003).\\
 \noindent [29] \ E.K. Lenzi, R.S. Mendes, L.C. Malacarne, L.R. da Silva, \ \emph{Physica A} {\bf 342}, \ 16 \ (2004).\\
 \noindent [30] \ V. Schw\"{a}mmle, E.M.F. Curado, F.D. Nobre, \ \emph{Eur. Phys. J. B}  {\bf 58}, 159 \ (2007).\\
 \noindent [31] \ V. Schw\"{a}mmle, F.D. Nobre, C. Tsallis, \ \emph{Eur. Phys. J. B}  {\bf 66}, \ 537 \ (2008).\\
 \noindent [32] \ F.D. Nobre, M.A. Rego-Monteiro, C. Tsallis, \ \emph{Phys. Rev. Lett.} {\bf 106}, \ 140601 \ (2011).\\
 \noindent [33] \ A.R. Plastino, C. Tsallis, \ \emph{J. Math. Phys.} {\bf 54}, \ 041505 \ (2013).\\
 \noindent [34] \  S. Abe, \ \emph{Europhys. Lett.} {\bf 90}, \ 50004 \ (2010).\\
 \noindent [35] \ B. Andresen, \ \emph{Europhys. Lett.} {\bf 92}, \ 40005 \ (2010).\\ 
 \noindent [36] \ S. Abe, \ \emph{Europhys. Lett.} {\bf 92}, \ 40006 \ (2010).\\
 \noindent [37] \ G.B. Bagci, T. Oikonomou, U. Tirnakli, \ \emph{Comment on ``Essential discreteness in \\
                                    \hspace*{8mm} generalised thermostatistics with non-logarithmic entropy" by S. Abe,}  {\sf arXiv:1006.1284}\\ 
 \noindent [38] \ J.P. Boon, J.F. Lutsko, \ \emph{Phys. Lett. A} {\bf 375}, \ 329 \ (2011).\\
 \noindent [39] \ J.F. Lutsko, J.P. Boon, \ \emph{Europhys. Lett.} {\bf 95}, \ 20006 \ (2011).\\ 
 \noindent [40] \ P. Quarati, M. Lissia, \ \emph{Emtropy} {\bf 15}, \ 4319 \ (2013).\\
 \noindent [41] \ A. Plastino, M.C. Rocca, \ \emph{Possible Divergences in Tsallis' Thermostatistics}, {\sf arXiv:1309.5645}\\  
 \noindent [42] \ N. Kalogeropoulos, \  \emph{QScience Connect}, \  {\bf 12} \ (2012).\\
 \noindent [43] \ N. Kalogeropoulos, \  \emph{QScience Connect}, \  {\bf 26} \ (2013).\\ 
 \noindent [44] \ N. Kalogeropoulos, \  \emph{J. Phys. Conf. Ser.} {\bf 410}, \ 012148 \ (2013).\\ 
 \noindent [45] \ N. Kalogeropoulos, \  \emph{Escort distributions and Tsallis entropy}, \ {\sf arXiv:1206.5127}\\
 \noindent [46] \ M. Gromov, \ \emph{Hyperbolic groups}, \  \ in \ \  \emph{Essays in group theory}, \ \ S. Gersten (Ed.),\\
                                \hspace*{8mm} MSRI Publ. {\bf 8}, \  Springer \ (1987).\\
 \noindent [47] \ P. Ontaneda, \ \emph{Pinched smooth hyperbolization} \ {\sf arXiv:1110.6374}\\
 \noindent [48] \ J. Cheeger, D.G. Ebin, \ \emph{Comparison Theorems in Riemannian Geometry}, \\
                     \hspace*{8mm} AMS \ Chelsea \  (1975).\\ 
 \noindent [49] \ H. Federer, \ \emph{Geometric Measure Theory}, \ Springer-Verlag, \ Berlin \ (1969).\\ 
 \noindent [50] \ F. Baldovin, M.-G. Moyano, A.-P. Majtey, A. Robledo, C. Tsallis,  \emph{Physica A} {\bf 340}, 205
                     \hspace*{8mm}  (2004).\\
 \noindent [51] \ G. Casati, C. Tsallis, F. Baldovin, \ \emph{Europhys. Lett.} {\bf 72}, \ 355 \ (2005).\\  
 \noindent [52] \ A. Pluchino, A. Rapisarda, C. Tsallis, \ \emph{Europhys. Lett.} {\bf 80}, \ 26002 \ (2007).\\
 \noindent [53] \ G. Lukes-Gerakopoulos, N. Voglis, C. Efthymiopoulos, \ \emph{Physica A} {\bf 387}, \ 1907 \ (2008).\\ 
 \noindent [54] \ A. Pluchino, A. Rapisarda, C. Tsallis, \ \emph{Physica A} {\bf 387}, \ 3121 \ (2008).\\
 \noindent [55] \ H.J. Hilhorst, G. Schehr, \ \emph{J. Stat. Mech.} {\bf P06003} \ (2007).\\
 \noindent [56] \ H.J. Hilhorst, \ \emph{Braz. J. Phys.} {\bf 39}, \ 371 \ (2009).\\ 
 \noindent [57] \ Ch. Antonopoulos, T. Bountis, V. Basios, \ \emph{Physica A} {\bf 390}, 3290 \ (2011).\\
 \noindent [58] \ L.-S. Young, \ \emph{J. Stat. Phys.} {\bf 108}, \ 733 \ (2002).\\
 \noindent [59] \ M. Gromov, V.D. Milman, \ \emph{Amer. J. Math.} {\bf 105}, \ 843 \ (1983).\\
 \noindent [60] \ C. Tsallis, M. Gell-Mann, Y. Sato, \ \emph{Proc. Nat. Acad. Sci.} {\bf 102}, \ 15377 \ (2005).\\
 \noindent [61] \ R. Hanel, S. Thurner,  \ \emph{Europhys. Lett.} {\bf 96}, \ 50003 \ (2011). \\
 \noindent [62]  \ N. Kalogeropoulos, \  \emph{Physica A} {\bf 391}, \ 3435 \ (2012).\\
 \noindent [63] \ E. Saksman, \ \emph{Ann. Acad. Sci. Fenn. Math.} {\bf 24}, \ 155 \ (1999).\\  
 \noindent [64] \ P. Assouad, \ \emph{Bull. Math. Soc. France} {\bf 111}, \ 429 \ (1983).\\
 \noindent [65] \ M. Gromov, \ \  \emph{Metric Structures for Riemannian and non-Riemannian spaces}, \\ 
                                 \hspace*{8mm} Birkh\"{a}user, \  Basel \ (1999).\\
 \noindent [66] \ R. Macias, C. Segovia, \ \emph{Adv. Math.} {\bf 33}, \ 257 \ (1979).\\
 \noindent [67] \ S. Abe, S. Martinez, F. Pennini, A. Plastino, \ \emph{Phys. Lett. A} {\bf 281}, \ 126 \ (2001).\\
 \noindent [68] \ G.L. Ferri, S. Martinez, A. Plastino, \ \emph{Physica A} {\bf 345}, \ 493 \ (2005).\\  
 \noindent [69] \ T. Wada, A.M. Scarfone, \ \emph{Phys. Lett. A} {\bf 335}, \ 351 \ (2005).\\
 \noindent [70] \ M. Nauenberg, \ \emph{Phys. Rev. E} {\bf 67}, \ 036114 \ (2003).\\
 \noindent [71] \ C. Tsallis, \ \emph{Phys. Rev. E} {\bf 69}, \ 038101 \ (2004).\\
 \noindent [72] \ M. Nauenberg, \ \emph{Reply to C. Tsallis's ``Comments on Critique of q-entropy for thermal \\
                               \hspace*{8mm} statistics by M. Nauenberg"}, \ {\sf arXiv:cond-mat/0305365}\\
 \noindent [73] \ A. Plastino, M.C. Rocca, \ \emph{Possible Divergences in Tsallis' Thermostatistics}, \ {\sf arXiv:1309.5645}\\ 
 \noindent [74] \ N. Kalogeropoulos, \ \emph{Ricci curvature, isoperimetry and Tsallis entropy}, \ \  (In preparation).\\

\end{document}